\documentstyle[12pt]{article}                    
\newfont{\ffont}{msym10}                          
\newcommand{\beq}{\begin{equation}}               
\newcommand{\eeq}{\end{equation}}                 
\newcommand{\bqry}{\begin{eqnarray}}              
\newcommand{\eqry}{\end{eqnarray}}                
\newcommand{\bqryn}{\begin{eqnarray*}}            
\newcommand{\eqryn}{\end{eqnarray*}}              
\newcommand{\preprint}[1]{\begin{table}[t]        
            \begin{flushright}                    
            \begin{large}{#1}\end{large}          
            \end{flushright}                      
            \end{table}}                          
\newcommand{\PD}[2]                               
    {\frac{\partial^{#2}}{\partial #1^{#2}}}      
\begin{document}
\preprint{LA-UR-97-3794rev}
\title{New Glueball-Meson Mass Relations}
\author{\\ M.M. Brisudova,\thanks{E-mail: BRISUDA@T5.LANL.GOV} \
L. Burakovsky,\thanks{E-mail: BURAKOV@T5.LANL.GOV} \ and \
T. Goldman\thanks{E-mail: GOLDMAN@T5.LANL.GOV}
\\  \\  Theoretical Division, MS B285 \\  Los Alamos National
Laboratory \\
Los Alamos, NM 87545, USA \\}
\date{\today}
\maketitle
\begin{abstract}
Using the ``glueball dominance'' picture of the mixing between
$q\bar{q}$ mesons of different hidden flavors, we establish new
glueball-meson mass relations which serve as a basis for glueball
spectral systematics. For the tensor glueball mass $2.3\pm 0.1$ GeV
used as an input parameter, these relations predict the following
glueball masses: $M(0^{++})\simeq 1.65\pm 0.05$ GeV, $M(1^{--})\simeq
3.2\pm 0.2$ GeV, $M(2^{-+})\simeq 2.95\pm 0.15$ GeV, $M(3^{--})\simeq
2.8\pm 0.15$ GeV. We briefly discuss the failure of such relations for the
pseudoscalar sector. Our results are consistent with (quasi)-linear Regge 
trajectories for glueballs with slope $\simeq 0.3\pm 0.1$ GeV$^{-2}.$
\end{abstract}
\bigskip
\newpage
\section{Introduction}
The existence of a gluon self-coupling in QCD suggests that, in
addition to the conventional $q\bar{q}$ states, there may be
non-$q\bar{q}$ mesons: bound states including gluons (glueballs and
$q\bar{q}g$ hybrids). However, the theoretical quidance on the
properties of unusual states is often contradictory, and models that
agree in the $q\bar{q}$ sector differ in their predictions about new
states. Moreover, the abundance of $q\bar{q}$ meson states in the 1-2
GeV region and glueball-quarkonium mixing makes the identification of
the would-be lightest non-$q\bar{q}$ mesons extremely difficult. To
date, no glueball state has been firmly established. 

Although the current situation with the identification of glueball
states is rather complicated, some progress has been made recently in
the $0^{++}$ scalar and $2^{++}$ tensor glueball sectors, where both
experimental and QCD lattice simulation results seem to converge
\cite{Land}. Recent lattice calculations predict the $0^{++}$ glueball
mass to be $1600\pm 100$ MeV \cite{Land,Bali,SVW,MP}. Accordingly,
there are two experimental candidates \cite{pdg}, $f_0(1500)$ and
$f_0(1710),$ in this mass range which cannot both fit into the scalar
meson nonet, and this may be considered as strong evidence for one of
these states being a scalar glueball (and the other being dominantly
$s\bar{s}$ scalar quarkonium).

In the tensor sector, the situation seems cleaner, though less well
established. Lattice simulations predict the $2^{ ++}$ glueball mass at
$2390\pm 120$ MeV \cite{MP,Peard}, and correspondingly, there are three
experimental candidates in this mass region \cite{pdg}, $f_J(
2220),\;J=2\;{\rm or}\;4,$ $f_2(2300)$ and $f_2(2340).$ The first
candidate is seen in $J/\psi \rightarrow \gamma +X$ transitions but not
in $\gamma \gamma $ production \cite{pdg}, while the remaining two are
observed in the OZI rule-forbidden process $\pi p\rightarrow \phi \phi
n$ \cite{pdg}, which favors the gluonium interpretation of all three
states.

Spin-1 glueballs are more complicated. Lattice studies on the vector
glueball are scarce and inconclusive \cite{Bali}, mainly because of the
difficulties in constructing the corresponding lattice
operators.\footnote{We thank W. Lee for this remark.} Various arguments
(e.g. \cite{Bali,HS}) suggest that the lowest lying $1^{--}$ glueball
has to consist of at least three constituent gluons.  Therefore, it is
heavier and more difficult to produce than the scalar and tensor
glueballs. On the other hand, once produced it should be easier to
identify since it can be expected to mix less with $q\bar{q}$
mesons.\footnote{The transitions between glueballs built of three
gluons and quarkonia are order $\alpha_S^{3/2}$, compared to order
$\alpha_S$ for the glueballs consisting of 2 gluons. If $\alpha_S < 1$
at the scale relevant for this transition, vector glueball mixing with
quarkonia is suppressed.}

In this paper we wish to undertake an attempt towards glueball spectral
systematics using mass relations. Our previous experience with mass
relations derived within different approaches to both light and heavy
mesons \cite{mes,BG1,BG2}, and to baryons \cite{bar}), shows that these
relations can be very successful. They typically hold with an accuracy
of a few percent, and often even 1\%. In order to relate the glueball
masses to the masses of known $q\bar{q}$ mesons, we need to identify
processes which are dominated by gluonic intermediate states. 

Such processes are, for example, OZI suppressed transitions~\cite{OZI}
between different hidden flavor states.  For this suppression to hold,
it has been shown that contributions from $q\bar{q}$ intermediate
states \cite{Lip} (even though not OZI suppressed, e.g., $\phi
\rightarrow K\bar{K}\rightarrow \rho \pi$) must (and do)
cancel~\cite{Lip2, GeI, HK}.  We assume this cancellation is
essentially complete and further assume that of all gluonic
intermediate states, the quark $q\bar{q}\leftrightarrow
q^{'}\bar{q}^{'}$ transition is dominated by the glueball with the
corresponding quantum numbers~\cite{Hou} which is closest in mass to
the mixing states. Under these basic assumptions, we relate the mass of
the glueball to the masses of the $q\bar{q}$ and $q^{'}\bar{q}^{'}$
mesons.

The paper is organized as follows: Section 2 relates quark mixing
amplitudes to the masses of physical mesons. In section 3, these
amplitudes are expressed in terms of glueball masses, and the new mass
relations are derived. We discuss the self-consistency of the
calculation and the results.  In section 4 we show that the glueball
masses we find are consistent with the expected gluoninc Regge
trajectories. The last section contains summary and conclusions.

\section{Meson mass squared matrix}
We start with the mass squared matrix in the (constituent) basis
$u\bar{u},$ $d\bar{d},$ $s\bar{s},$ modified by the inclusion of the
quark mixing amplitudes. (In the following, the symbol for the meson
stands for its mass, unless otherwise specified, and we use the
notations $\rho ,$ $K^\ast ,$ $\omega $ and $\phi $ for the isovector,
isodoublet, and two isoscalar states, respectively, of a meson nonet of
any spin.) Let
\beq
{\cal M}^2=\left(
\begin{array}{ccc}
\rho ^2+A & A & A^{'} \\
A & \rho ^2+A & A^{'} \\
A^{'} & A^{'} & 2K^{\ast 2}-\rho ^2+A^{''}
\end{array}
\right) ,
\eeq
where the quark mixing amplitudes $A$, $A^{'}$, $A^{''}$ are assumed to
be isospin- but not SU(3)$_f$-symmetric, i.e., $A_{uu}=A_{ud}=
A_{dd}\equiv A,$ $A_{us}=A_{ds}\equiv A^{'},$ and $A_{ss}\equiv
A^{''}.$ Transforming the mass squared matrix (1) to the (Gell-Mann)
basis $(u\bar{u}-d\bar{d})/\sqrt{2},$
$(u\bar{u}+d\bar{d}-2s\bar{s})/\sqrt{6},$
$(u\bar{u}+d\bar{d}+s\bar{s})/\sqrt{3},$ one obtains
\beq
{\cal M}^2=\left( \!
\begin{array}{ccc}
\rho ^2 & 0 & 0 \\
0 & \!\frac{1}{3}(4K^{\ast 2}\!-\!\rho
^2)\!+\!\frac{2}{3}(A\!-\!2A^{'}\!+\!A^{
''})\! & \!-\frac{2\sqrt{2}}{3}(K^{\ast 2}\!-\!\rho
^2)\!+\!\frac{\sqrt{2}}{3}
(2A\!-\!A^{'}\!-\!A^{''}) \\
0 & \!-\frac{2\sqrt{2}}{3}(K^{\ast 2}\!-\!\rho
^2)\!+\!\frac{\sqrt{2}}{3}(2A\!-
\!A^{'}\!-\!A^{''}) & \frac{1}{3}(2K^{\ast 2}\!+\!\rho
^2)\!+\!\frac{1}{3}(4A
\!+\!4A^{'}\!+\!A^{''})
\end{array}
\!\right) \!,
\eeq
which shows that the mixing amplitudes contribute only to the subspace
spanned by the isoscalar (self-conjugate) states (as we have not
included the small effect of isospin symmetry breaking). In the
following, we restrict ourselves to this subspace alone, namely the
lower right $2\times 2$ block ($M^2$ matrix) of Eq. (2).

The masses of the physical isoscalar states ($\omega $ and $\phi $) are
given by diagonalizing the $M^2$ matrix,
\beq
M^2=\left(
\begin{array}{cc}
\phi ^2 & 0 \\
0 & \omega ^2
\end{array}
\right) ,
\eeq
and may be obtained from the invariance of the trace and determinant
under a unitary transformation, such as rotation in the flavor space,
viz.,
\beq
{\rm Tr}\;\!M^2=2K^{\ast 2}+2A+A^{''}=\omega ^2+\phi ^2,
\eeq
\beq
{\rm Det}\;\!M^2=\left( 2K^{\ast 2}+\rho ^2+A^{''}\right) \left( \rho
^2+2A
\right) -2A^{'2}=\omega ^2\phi ^2.
\eeq

The transition amplitudes are proportional to unknown matrix elements
of the (unspecified) effective Hamiltonian. In order to reduce the
number of parameters, we try to relate the amplitudes $A,A^{'}$ and
$A^{''}$. It is plausible to assume that
\beq
AA^{''}\simeq A^{'2}.
\eeq
This relation is rigorous for the pseudoscalar mesons \cite{BG3}, and holds
for the parametrization of the two-gluon-induced transition amplitude in the
form
\beq
A_{ij}=\frac{\Lambda }{M_iM_j},
\eeq
where $\Lambda $ is an SU(3)-invariant parameter and $M_i,M_j$ are the
constituent quark masses \cite{GI}. Here we propose the validity of the
relation Eq. (6) for a quark mixing amplitude with an arbitrary number of
gluons. 

In accordance with Eq. (6), we introduce a parameter $r:$
\beq
A^{'}=Ar,\;\;\;A^{''}=Ar^2,
\eeq
Since from Eq. (7) $A>A^{'}>A^{''}$, one can expect $r\leq 1$. 

Eqs. (4), (5) can be rewritten in terms of $r$ and $A$ as follows:
\beq
\omega ^2+\phi ^2=2K^{\ast 2}+A(2+r^2),
\eeq
\beq
\left[ 4K^{\ast 2}-(2+r^2)\omega ^2-(2-r^2)\rho ^2\right] \left[
(2+r^2)\phi ^
2+(2-r^2)\rho ^2-4K^{\ast 2}\right] =8r^2\!\left( K^{\ast 2}\!-\rho
^2\right) ^
2\!\!.
\eeq
Eq. (10) is a modified version of Schwinger's quartic mass formula
\cite{Sch}, in which annihilation effects are taken into account. It is
interesting to note that in the case of SU(3)-invariant quark mixing
amplitudes, (i.e.  $r=1$) it reduces to Schwinger's original relation,
{\em independent of the value of} $A$.  

Using Eqs.~(4), (5), (8), $A$ and $r$ can be expressed in terms of the
meson masses~\cite{Sca}:
\bqry
A & = & \frac{1}{4}\;\frac{(\omega ^2-\rho ^2)(\phi ^2-\rho
^2)}{K^{\ast 2}-\rho ^2}, \\
r^2 & = & 2\;\frac{(\phi ^2+\rho ^2-2K^{\ast 2})(2K^{\ast 2}-\rho ^2-\omega
^2)}{(\phi ^2-\rho ^2)(\omega ^2-\rho ^2)}.
\eqry

Note that $A$ is small due to the near mass-degeneracy of $\rho $ and
$\omega $ states which is related to the near-ideal mixing. This
smallness is a confirmation of OZI rule. Since both the denominator and
numerator of Eq. (12) contain nearly vanishing factors (the $\rho $ and
$\omega $ numerator factor of Eq. (11) and a factor which would vanish
if the  Gell-Mann--Okubo relation were exact, respectively), any small
change in the mass values induces a large change in $r^2.$ (For
example, a one per cent change in the $K^\ast $ mass makes r
imaginary.) Conversely, the masses derived from the relations below are
relatively insensitive to the value of $r.$ To determine the glueball
masses by those relations, we choose a set of meson masses which give
$r \leq 1$. Masses we use as input are given in Table 
I~\cite{pdg,DM,rho-t,rho-e}. \\
\begin{center}
\begin{tabular}{|c|c|c|c|c|} \hline
 $J^{PC}$ & $I=1$ & $I=1/2$ & $I=0,\;\omega $ & $I=0,\;\phi $  \\ \hline
 $0^{-+}$ & 0.1373 & 0.4957 & 0.9578 & 0.5475  \\ \hline
 $0^{++}$ & 1.318  & 1.429  & 0.98/1.1 &  1.5  \\ \hline
 $1^{--}$ & 0.760  & 0.8961 & 0.8919 & 1.0194  \\ \hline
 $2^{++}$ & 1.318  & 1.429  & 1.275  &  1.525  \\ \hline
 $2^{-+}$ &  1.67  &  1.78  &  1.65  &  1.88   \\ \hline
 $3^{--}$ &  1.69  &  1.78  &  1.67  &  1.86   \\ \hline
\end{tabular}
\end{center}
{\bf Table I.} Meson masses (in GeV). $I$ stands for isospin, and
$\omega ,\phi $ indicate the isoscalar mostly singlet and octet states,
respectively.
 \\
\section{Glueball-meson mass relations}
In this section we derive the meson-glueball mass relations.  

The OZI suppression rule may be interpreted in terms of the Feynman box
graph connecting the annihilation of $q\bar{q}$ of one flavor and the
pair creation of $q^{'}\bar{q}^{'}$ of another flavor, plus all gluonic
and quark loop dressings thereof. To obtain a physical interpretation,
it is convenient to consider the various time orderings of these graphs
in the overall rest frame of the annihilating pair. Here we see that
graphs with the form of an overlapping double hairpin reflect
multi-meson $(q\bar{q}$ meson) intermediate states, while others
reflect essentially purely gluonic (ignoring fully closed quark loops
such as would be eliminated by quenching in lattice calculations)
intermediate states. The results of ref.  \cite{GeI} suggest that the
former strongly cancel, and we extend this to mean their contribution
is entirely negligible. For the latter, which describe the usual
interpretation of OZI suppression, it is natural to attempt to estimate
the strength by saturating with pure glue (glueball) resonances. 

Let us therefore assume that the $q\bar{q}\leftrightarrow
q^{'}\bar{q}^{'}$ transition proceeds via gluonic intermediate states,
viz.,
\beq
A_{qq^{'}}=\sum _i\frac{\langle
q\bar{q}|H|i\rangle
\langle i|H|q^{'}\bar{q}^{'}\rangle
}{M^2-
M^2(i)},
\eeq
where $H$ is the effective transition Hamiltonian, $|i\rangle $ is a
complete set of gluonic states, $M^2(i)$ is mass squared of the
intermediate state, and $M^2$ is mass squared of the initial (and
final) state. We now further assume that the sum (13) is saturated by
the lowest lying glueball with the corresponding quantum numbers.
Therefore, for $q,q^{'}=n(=u,d)$,
\beq
\Big| A\Big| \equiv \Big| A_{nn}\Big| \simeq \left| \frac{f^2(\omega _n^2)}{
\omega _n^2-G^2}\right| ,\;\;\;f(\omega _n^2)\equiv \langle G|H
|n\bar{n}\rangle \Big| _{q^\mu q_\mu =\omega _n^2},
\eeq
and for $q,q^{'}=s,$
\beq
\Big| Ar^2\Big| \equiv \Big| A_{ss}\Big| \simeq \left| \frac{f^2(\omega
_s^2)}{\omega _s^2-G^2}\right| ,\;\;\;f(\omega _s^2)\equiv \langle
G|H |s\bar{s}\rangle \Big| _{q^\mu q_\mu =\omega _s^2},
\eeq
where $\omega _n^2=\rho ^2$ and $\omega _s^2=2K^{\ast 2}-\rho ^2$ are
the masses squared of pure $q\bar{q}$ counterparts of the physical
$\omega $ and $\phi$ states, and $G^2$ is the corresponding glueball
mass squared. Note that we include the lowest lying glueball only. Even
though transitions via excited glueballs are suppressed by both the
numerator and denominator of $(13)$, it is of course not a priori clear
that the sum can be well approximated by the first term only.

To proceed further, it is necessary to have some information regarding
the functions $f(\omega ^2).$ If one considered an analogous situation
in a solved theory (e.g., nonrelativistic QED for the bound states of
lepton-antilepton pairs of various flavors), then one would expect the
magnitude of an analog of $f(\omega ^2)$ to vary markedly with both
orbital and radial quantum numbers. We assume, however, that the
product of $f(\omega _n^2)$ and $f(\omega _s^2)$ in QCD is a constant
approximately independent of the quantum numbers of a meson nonet,
viz.,
\beq
f(\omega _n^2)f(\omega _s^2)\approx
{\rm const.}
\eeq
(This assumption is more general than the one used in ref. \cite{HK}
where $f(\omega ^2)$ itself is assumed to be independent of $\omega
^2$.) We will show that this counter-intuitive assumption is justified
{\it ex post facto} by the results obtained, and we will check it for
self-consistency below.  

Representing now the product $A\cdot Ar^2$ in two ways, viz., from
Eqs.  (11), (12), and (14), (15), and using (16), one obtains a set of
mass relations:
$$\frac{(T^2-a_2^2)(T^2+a_2^2-2K_2^{\ast 2})}{(V^2-\rho ^2)(V^2+\rho
^2-2K^{\ast 2})}\approx \left( \frac{K_2^{\ast 2}-a_2^2}{K^{\ast2}-\rho
^2}\right) ^2$$
\beq
\times \frac{(\phi ^2+\rho ^2-2K^{\ast 2})(2K^{\ast 2}-\rho ^2-\omega
^2)(\phi ^2-\rho ^2)(\omega ^2-\rho ^2)}{(f_2^{'2}+a_2^2-2K_2^{\ast
2})(2K_2^{\ast 2}-a_2^2-f_2^2)(f_2^{'2}-a_2^2)(f_2^2-a_2^2)},
\eeq
$$\frac{(T^2-a_2^2)(T^2+a_2^2-2K_2^{\ast 2})}{(V_3^2-\rho _3^2)(V_3^2+\rho
_3^
2-2K_3^{\ast 2})}\approx \left( \frac{K_2^{\ast 2}-a_2^2}{K_3^{\ast 2}-\rho
_
3^2}\right) ^2$$
\beq
\times \frac{(\phi _3^2+\rho _3^2-2K_3^{\ast 2})(2K_3^{\ast 2}-\rho _3^2-
\omega _3^2)(\phi _3^2-\rho _3^2)(\omega _3^2-\rho
_3^2)}{(f_2^{'2}+a_2^2-2K_
2^{\ast 2})(2K_2^{\ast 2}-a_2^2-f_2^2)(f_2^{'2}-a_2^2)(f_2^2-a_2^2)},\;\;\;
{\rm etc.,}
\eeq
where $V,T,V_3,\ldots $ are the masses of the vector, tensor,
$3^{--},\ldots $ glueballs.\footnote{In Tables II and III below,
$T^{'}$ stands for the mass of the $2^{-+}$ glueball.} 

It is apparent from these relations that, if one of the glueball masses
is chosen as an input parameter, the masses of other glueballs can be
predicted, provided the corresponding $(q\bar{q})$ meson masses are
known. We choose the mass of the tensor glueball as such an input
parameter, $T=2.3\pm 0.1$ GeV, and calculate the masses of the
$0^{-+},$ $0^{++},$ $1^{ --},$ $2^{-+}$ and $3^{--}$ glueballs. (We do
not calculate the masses of any other glueball states, since the
$q\bar{q}$ assignments of the corresponding meson multiplets are not
established so far.) Our results are presented in Table II. \\
\begin{center}
\begin{tabular}{|c|c|c|c|c|c|c|c|} \hline
$T$ & $V$ & $T^{'}$ & $V_3$ & $PS$ & $S_1$ & $S_2$  \\ \hline
 2.2 & 2.934 & 2.805 & 2.627 & 0.877 & 1.587 & 1.624  \\ \hline
 2.3 & 3.143 & 2.937 & 2.740 & 0.916 & 1.606 & 1.646  \\ \hline
 2.4 & 3.347 & 3.069 & 2.852 & 0.956 & 1.623 & 1.670  \\ \hline
\end{tabular}
\end{center}
{\bf Table II.} Predictions for the glueball masses (in GeV) for three
input values of the tensor glueball mass: 2.2 GeV, 2.3 GeV, 2.4 GeV.
\\

In calculating the glueball masses in Table II, we used the meson
masses given in Table I. Also, in calculating the value of the
scalar glueball mass, mass degeneracy of the corresponding isovector
and isodoublet states of the tensor and scalar meson nonets was
assumed, in agreement with ref.  \cite{P-wave}: $a_0=a_2,$ $K_0^\ast
=K_2^\ast .$\footnote{The latter is in agreement with data \cite{pdg};
the experimental candidate which satisfies the former $(a_0=1322\pm 30$
MeV) was seen by LASS \cite{LASS} and GAMS \cite{GAMS}. As for the
candidate $a_0(1450)$ contained in \cite{pdg}, we note that, as we have
checked, the results are not very sensitive to the mass of the scalar
isovector state: For $a_0\simeq 1.4\pm 0.25$ GeV, relations of both the
type in Eqs. (17) and (18) and in Eqs. (19) and (20) below, predict the
scalar glueball mass in the range $\simeq 1.65\pm 0.05$ GeV.} The mass
of the isoscalar mostly octet state was taken to be 1.5 GeV
\cite{pdg,scalar2}, and two cases were considered:  The mass of the
isoscalar mostly singlet state was taken to be 0.98 GeV \cite{pdg} or
1.1 GeV \cite{scalar2} (for which the values of $r,$ as calculated from
(12), are, respectively, 0.82 and 0.90), and the results presented in
Table II under $S_1$ and $S_2,$ respectively.

\subsection*{Approximation $r\approx 1$}

Let us also consider the appproximation of flavor-independent quark
annihilation amplitudes. It then follows from Eq. (4) (with $A^{''}=A)$
that $A=(\phi ^2+\omega ^2-2K^{\ast 2})/3,$ and Eqs. (17), (18) are
replaced by
\bqry
\frac{(T^2-a_2^2)(T^2+a_2^2-2K_2^{\ast 2})}{(V^2-\rho ^2)(V^2+\rho
^2-2K^{\ast
2})} & \approx  & \left( \frac{\phi ^2+\omega ^2-2K^{\ast 2}}{f_2^2+
f_2^{'2}-2K_2^{\ast 2}}\right) ^2, \\
\frac{(T^2-a_2^2)(T^2+a_2^2-2K_2^{\ast 2})}{(V_3^2-\rho _3^2)(V_3^2+\rho
_3^2-
2K_3^{\ast 2})} & \approx  & \left( \frac{\phi _3^2+\omega _3^2-2K_3^{\ast
2}}{f_2^2+f_2^{'2}-2K_2^{\ast 2}}\right) ^2,\;\;\;{\rm etc.}
\eqry
Glueball masses calculated from  Eqs. (19), (20) are presented
in Table III. \\
\begin{center}
\begin{tabular}{|c|c|c|c|c|c|c|c|} \hline
$T$ & $V$ & $T^{'}$ & $V_3$ & $PS$ & $S_1$ & $S_2$  \\ \hline
 2.2 & 3.016 & 2.794 & 2.714 & 0.919 & 1.601 & 1.646  \\ \hline
 2.3 & 3.232 & 2.925 & 2.836 & 0.952 & 1.619 & 1.672  \\ \hline
 2.4 & 3.443 & 3.056 & 2.958 & 0.985 & 1.638 & 1.699  \\ \hline
\end{tabular}
\end{center}
{\bf Table III.} The same as in Table II, according to mass relations
of the type given by Eqs. (19),(20).
\\

\noindent Comparison of the results given in Tables II and III shows
that they are not very sensitive to the precise values of $r,$ except
perhaps for the $1^{--}$ glueball.

The glueball masses depend only weakly on isovector masses. In order to
demonstrate this point, let us rewrite, e.g., Eq. (19) which
constitutes a quadratic equation for $V^2$, and expand it as follows:
\beq
V^2 = K^{\ast 2} + \sqrt{\Delta_T \over{\Delta_V}}
\left( T^2 - K_2^{\ast 2} \right)
-{1 \over{2}}\sqrt{\Delta_T \over{\Delta_V}}
{(K_2^{\ast 2}-a_2^2)^2 \over{ (T^2 - K_2^{\ast 2})}}
+{1 \over{2}}\sqrt{\Delta_V \over{\Delta_T}}
{(K^{\ast 2}-\rho^2)^2 \over{ (T^2 - K_2^{\ast 2})}} + \cdots ,
\eeq
where $\Delta_V =({\phi ^2+\omega ^2-2K^{\ast 2}})^2$ and
$\Delta_T=({f_2^2+ f_2^{'2}-2K_2^{\ast 2}})^2$. Note that each of 
these quantities describes violation of a Gell-Mann-Okubo relation, 
and so is not large; nonetheless, the values vary from multiplet to 
multiplet. To the extent that we obtain self-consistency and agreement
with experiment, this suggests that we have indeed identified the
origin of OZI-violating contributions in terms of glueball intermediate
states. (In deriving this relation we used the fact that the
$SU(3)$ violating terms ($K^{\ast 2}-\rho^2$, etc.) are small.)

	It is also important to recognize that in the Gell-Mann-Okubo limit, 
the ratio of $\Delta$'s in eqn. (21) is undefinied, so the results of our 
analysis are highly sensitive to the input meson masses which produce small 
violations of the Gell-Mann-Okubo relations.

\subsection*{Consistency check} 

We now return to check the consistency of Eq. (16). Using the following
glueball masses, (in GeV) $S=1.61,$ $V=3.2,$ $T=2.3,$ $V_3=2.8,$
$T^{'}=2.95,$ on the basis of the results presented in Tables II, III,
we calculate the values of $f(\omega _n^2),$ $f(\omega _s^2),$ with the
help of Eqs. (11), (12), (14) and (15), and the product $f(\omega
_n^2)f(\omega _s^2)$ for the five multiplets. (We use $f_0^{'}=980$ MeV
for the scalar meson nonet). The results are shown in Table IV. \\
\begin{center}
\begin{tabular}{|c|c|c|c|c|c|c|c|} \hline
$J^{PC}$ & $f(\omega _n^2),$ GeV$^2$ & $f(\omega _s^2),$ GeV$^2$ &
$f(\omega _n^2)f(\omega _s^2),$ GeV$^4$  \\ \hline
 $0^{++}$ & 0.528$ \pm $0.04 & 0.232 $\pm $0.08 & 0.122 $\pm $ 0.05 \\ \hline
 $1^{--}$ & 0.409 $\pm $ 0.05 & 0.303 $\pm $0.04 & 0.124 $\pm $ 0.03 \\ \hline
 $2^{++}$ & 0.437 $\pm $0.06 & 0.273 $\pm $0.04 & 0.119 $\pm $ 0.03 \\ \hline
 $2^{-+}$ & 0.439 $\pm $0.07 & 0.275 $\pm $ 0.05 & 0.121 $\pm $ 0.04 \\ \hline
 $3^{--}$ & 0.402 $\pm $0.07 & 0.319 $\pm $0.06 & 0.128 $\pm $ 0.05 \\ \hline
\end{tabular}
\end{center}
{\bf Table IV.} The values of $f(\omega _n^2),$ $f(\omega _s^2),$
$f(\omega _n^2)f(\omega _s^2),$ 
for the five meson multiplets. Error estimates were computed from the
variations induced by the range of mass values of $T.$
\\

Comparison of the results for $f(\omega _n^2)f(\omega _s^2)$ shows that
they are consistent with Eq.~(16), up to $\sim 7$\% accuracy, which is
in qualitative agreement with the accuracy of the values predicted for
the glueball masses, (e.g., $3.2\pm 0.2$ GeV for the vector glueball is
$\sim 6.5$\% accuracy). It is interesting to note that, in Table IV,
both $f(\omega _n^2)$ and $f(\omega _s^2)$ are in even closer agreement
for the particular glueball pairs: $(1^{--},3^{--})$ and
$(2^{++},2^{-+}).$)

Moreover, the assumption (16) seems to be justified for radial
excitations as well. We calculated the value of $ f(\omega
_n^2)f(\omega _s^2)$ for 2 $^3S_0$ multiplet with $\eta(1295)$,
$\pi(1350)$, $K(1430)$ and $\eta(1490)$ as input masses. The result is
again in remarkable agreement with the values given in Table IV.
Specifically, we obtained $ f(\omega _n^2)= 0.389$, $f(\omega _s^2)=
0.334$ and $f(\omega _n^2)f(\omega _s^2)=0.130$. The origin of the
validity of (16) remains a mystery at present.

\section{Further implications}

Let us examine our results from another point of view.  We have found
that the vector and spin-3 glueballs have masses around 3 GeV.  Can
this fact find its simple explanation in, e.g., QCD phenomenology? The
answer is positive. Both states are composed of three constituent
gluons, and a naive scaling from the two-gluon $2^{++}$ glueball to the
3-gluon case gives $M(3g)\simeq 1.5\;\!M(2g)\simeq 3.3$ GeV, with
$M(2g)\simeq 2.2$ GeV.  Also, the original constituent gluon model
predicts $M(1^{--})/M(2^{++}) \simeq M(3^{--})/M(2^{++})\simeq 1.5$
\cite{HS}. Note that the value of the vector glueball mass obtained in
this paper is consistent with the Brodsky-Lepage-Tuan domain \cite{BLT}
\beq
\Big| M(1^{--})-M(J/\psi )\Big| <80\;{\rm MeV,}
\eeq
obtained from the ratio of the measured widths: $\Gamma (\psi
^{'}\rightarrow V\rightarrow \rho \pi )/\Gamma (J/\psi \rightarrow
V\rightarrow \rho \pi ).$

The value of the pseudoscalar glueball mass obtained here, $\sim
0.93\pm 0.05$ GeV, however, is inconsistent with the lattice result
\cite{MP,Peard} $2490\pm 140$ MeV. A possible explanation for this is
that we have not included instanton effects. Instanton effects  are
irrelevant for all other channels, but they may be important for
pseudoscalars (and possibly scalars). Indeed, as has been shown
separately \cite{BG3}, non-instanton annihilation effects alone cannot
provide the mass splitting of $\sim 500$ MeV required to reproduce the
physical pseudoscalar meson spectrum. The use of the pseudoscalar
glueball mass $M(0^{-+})\simeq M(2^{++})=2.2-2.4$ GeV in relations of
the type (19), (20) leads to $\eta ^{'2}+\eta ^2-2K^2=0.08-0.09$
GeV$^2,$ in contrast to the 0.73 GeV$^2$ required by pseudoscalar meson
spectroscopy. The remaining $\sim 0.65$ GeV$^2$ would then be expected
to reflect the contribution of instantons \cite{BG3}, so the
introduction of instanton effects which are known to be strong in this
sector, in addition to those of gluons, changes the situation
drastically.

On the other hand, the gluon annihilation effects seem to be sufficient
to reproduce the scalar meson spectrum. The value obtained for the
scalar glueball mass, $\sim 1650\pm 50$ MeV, is in agreement with
lattice results $1600\pm 100$ MeV \cite{Bali,SVW,MP}. We note that the
mass predicted for the $2^{-+}$ glueball, $\simeq 2.95\pm 0.15$ GeV, is
also in agreement with lattice QCD results: $3.07\pm 0.15$ GeV
\cite{Peard} and $\simeq 3$ GeV \cite{Bali}. Finally, we note that QCD
sum rules predict $M(0^{++})=1.5\pm 0.2$ GeV, $M(2^{++})\simeq 2.0\pm
0.1$ GeV, $M(0^{-+})\simeq 2.05\pm 0.2$ GeV, and find the $3g$-glueball
mass to be $\simeq 3.1$ GeV \cite{Nar}.

All the glueball masses calculated above, except for the scalar one,
are much higher than those of the corresponding mesons, and therefore
cannot appreciably mix with the latter.\footnote{A mixing of the
quarkonia with the four-quark states of the same quantum numbers which
may have comparable masses, although not precluded, is not treated
here. Although we can no more justify this than any others do, we
suspect that justification may be related to the apparently consistent
neglect here of multi-($q\bar{q}$-)meson intermediate states in the OZI
violating mixing amplitudes.} Thus, besides the pseudoscalar glueball,
only the scalar glueball may be expected to mix considerably with the
scalar isoscalar states, since it lies in a mass range spanned by the
latter; for any other nonet the use of the $2\times 2$ mass matrix (3)
is justified. However, even for the scalar glueball, the mass shift
produced by the mixing with quarkonia need not necessarily be large. In
a model considered in ref. \cite{Wei}, for example, the bare glueball
mass of 1635 MeV is shifted up to 1710 MeV, and the bare $s\bar{ s}$
mass of 1516 MeV is shifted down to 1500 MeV, both modest effects. We
also note that relations of the type in Eqs. (17), (18) can be also
used to predict masses of problematic quarkonia; e.g., the mass of the
2 $^3S_1$ isodoublet state.

\subsection*{Glueball Regge trajectories}
Finally, we briefly consider the question of the glueball Regge
trajectories.  A knowledge of these trajectories may be useful for
determining  masses of glueballs with exotic quantum numbers (e.g.,
$0^{+-},$ $0^{--},$ $1^{-+},$ etc.), for which no $q\bar{q}$
counterparts exist so that relations of the type given by Eqs. (17) and
(18) or (19) and (20) therefore cannot be applied.

It is widely believed that the tensor glueball is the first particle
lying on the (quasi)-linear pomeron Regge trajectory \cite{Lan},
$\alpha (t)=\alpha _{\rm P}(0)+\alpha ^{'}_{\rm P}\cdot t.$ The
parameters have been extracted from experimental data on diffractive
deep-inelastic scattering: $\alpha _{\rm P}(0)=1.07\pm 0.03,$ $\alpha
^{'}_{ \rm P}=0.25\pm 0.01$ GeV$^{-2}$ \cite{KTM}, or $\alpha _{\rm
P}(0)= 1.086,$ $\alpha ^{'}_{\rm P}=0.25$ GeV$^{-2}$ \cite{DL}. Higher
values of $\alpha ^{'}_{\rm P}$ have been considered in the literature
(in GeV$^{-2}):$ $\alpha ^{'}_{\rm P}=0.3$ \cite{Dub}, 0.3-0.36
\cite{Volk}, 0.311 \cite{Sim}, 0.32-0.46 \cite{DLM}. These differences
may be reconciled if one assumes non-linearity of trajectories as
suggested theoretically in refs.~\cite{BuKo} and \cite{BuChiKo},
(taking into account the fact that the slopes have been extracted from
data in differing momentum transfer regions,) which may have been
experimentally observed at CERN~\cite{PomEx}.

If we take the glueball masses from Table III and ignore the difference
in intercepts of parity partner trajectories, then under the assumption
of a common linear trajectory for the $2^{++}$ and $3^{--}$ glueballs,
we obtain
\beq
\alpha ^{'}_G = \frac{1}{M^2(3^{--})-M^2(2^{++})} = 0.37 \pm 0.04
\eeq
where the error is taken from the variation induced by the range of
input masses for the tensor glueball. (The result from Table II is 25\%
larger.) Similar results may be obtained from other combinations. 

Our results, therefore, are consistent with the standard expectation
that the glueballs populate (quasi)-linear Regge trajectories with
slope $\simeq 0.3 \pm 0.1$ GeV$^{-2},$ in agreement with
refs.~\cite{KTM,DL,Volk,Sim,DLM}. It should be kept in mind, however,
that the power of this demonstration is limited by the difficulty of
examining the entire space of meson masses consistent with $r=1$. So
far, we have only studied the single point in that space defined by
Table I.

\section{Summary and Conclusions.}
In this paper we used OZI suppressed processes in the isoscalar sectors
to find new glueball-meson mass relations. First, motivated by a
parametrization of the two-gluon amplitude in terms of quark masses and
a hadronic scale $\Lambda$, we assumed that the transition amplitudes
for quark mixing can be related via a parameter $r$, which can be
expected to satisfy $r\leq 1$. Since the masses turn out to be
insensitive to the exact value of $r$, $r$ cannot be determined from
data, and we fixed the relevant meson masses to give $r\leq 1$. We then
assumed cancellation of two step hadronic contributions and glueball
dominance. Finally, we assumed that matrix elements of the (in practice
unknown) interaction Hamiltonian between $q\bar{ q}$ and glueball with
the corresponding quantum numbers are independent of the quantum
numbers of the meson nonet. Under these assumptions, we established new
glueball-meson mass relations, and used them to predict the glueball
masses. With the tensor glueball mass $2.2\pm 0.1$ GeV chosen as an
input parameter in these relations, we  obtained the following glueball
masses:
\bqryn
M(0^{++}) & \simeq  & 1.65\pm 0.05\;{\rm GeV,} \\
M(1^{--}) & \simeq  & 3.2\pm 0.2\;{\rm GeV,} \\
M(2^{-+}) & \simeq  & 2.95\pm 0.15\;{\rm GeV,} \\
M(3^{--}) & \simeq  & 2.8\pm 0.15\;{\rm GeV.}
\eqryn
We have shown that these glueball masses are consistent with
(quasi)-linear Regge trajectories, with slope $\simeq 0.3 \pm 0.1$
GeV$^{-2}.$ Our calculation is self-consistent, in the sense that the
results have not led to any contradictions with our assumptions.

Finally, we should comment on the unnatural, but apparently
self-consistent, assumption regarding the behavior of the annihilation
transition amplitudes ($f$), namely the state independence of their
light-quark-strange-quark product, which does not seem to depend upon
orbital or even radial quantum numbers of the quarks involved. If the
glueball mass spectrum we derive is experimentally confirmed,
supporting our plethora of assumptions, it will be necessary to find a
physical interpretation of this regularity. Here we merely note that
while perturbative analyses would suggest a strong variation, they
depend on a specific behavior of the t-channel segment of the quark
propagator involved in the annihilation.  However, if this exchange is
also (quark-) Reggeized~\cite{Nus}, then the amplitude could be
expected to be dominated by the trajectory intercept, i.e., at the
minimum momentum transfer possible. This quantity is indeed multiplet
independent, so long as the trajectories involved are degenerate
(rather than split by parity, or daughters, for example). It would be
intriguing to confirm such a consistency between Regge trajectories for
quarks and color singlet hadrons.

\section*{Acknowledgments}
One of us (L.B.) wishes to thank Weonjong Lee for very valuable
discussions during the preparation of this work.

\end{document}